\PassOptionsToPackage{expansion=false}{microtype}
\documentclass{ceurart}

\usepackage{graphicx}
\usepackage{xspace}

\newcommand{\scratch}{\texttt{Scratch}\xspace}
\newcommand{\cb}{\textit{custom block}\xspace}
\newcommand{\cbs}{\textit{custom blocks}\xspace}

\newcommand{\paperurl}{\url{https://figshare.com/s/4777dacbe5ef9c0f0dad}}

\begin{document}

\copyrightyear{2026}
\copyrightclause{Copyright for this paper by its authors. Use permitted
  under Creative Commons License Attribution 4.0 International (CC BY 4.0).}

\conference{BENEVOL 2026: The 25th Belgium-Netherlands Software Evolution
  Workshop, November 26--27, 2026, Madrid, Spain}

\title{Code Cloning and Procedural Abstraction in Scratch:
  A Large-Scale Study of Co-Existing Reuse Practices}

\author[1]{Ricardo Hidalgo-Aragon}[email=r.hidalgoa.2024@alumnos.urjc.es,]
\cormark[1]
\author[2]{Jes\'us Moreno-Le\'on}[email=jmoreno@us.es,]
\author[3]{Alejandro Echaniz}[email=10alex.echa@gmail.com]
\author[1]{David Rold\'an-\'Alvarez}[email=david.roldan@urjc.es,]
\author[1]{Jes\'us M. Gonz\'alez-Barahona}[email=jgb@urjc.es,]
\author[1]{Gregorio Robles}[email=grex@urjc.es,]

\address[1]{Universidad Rey Juan Carlos, Madrid, Spain}
\address[2]{Universidad de Sevilla, Sevilla, Spain}
\address[3]{Independent Researcher, Spain}

\cortext[1]{Corresponding author.}

\begin{abstract}
Code cloning is a long-standing software-evolution concern: duplicated implementations diverge as software changes. In block-based programming,
however, duplication is an everyday reuse mechanism, and \scratch facilitates a copy-and-paste culture that may hinder abstraction. We 
study this relation in 2.04 million public \scratch projects. We compare cloned sequences with existing abstractions and examine how their use varies across Computational Thinking~(CT) score levels.

Clones occur in 6.2\% of projects and \cbs in 11.9\%. Of 436,198 detected clone 
instances, 16.1\% have a normalized sequence matching a \cb in the same project 
and 17.8\% match a \cb only outside it. At the pattern level, 83 of the 100 most 
frequent clone patterns have counterparts implementable as custom blocks elsewhere 
in the corpus. Of the remaining 17 patterns, 15 are amenable to custom-block 
extraction, while in two cases the benefit of abstraction is ambiguous; no class 
of frequent clone resists abstraction in principle. Yet the 
abstractions that do exist are often shallow: 53.0\% of \cbs are parameter-free.

Increased use of abstraction does not coincide with less cloning. Both become more prevalent at higher CT-score levels, and their combined use dominates the highest band. These findings are consistent with cloning and procedural abstraction 
being co-existing reuse practices rather than mutually exclusive stages. Closing that 
gap requires both instruction on when to abstract, not only how, and platform support 
for refactoring.
\end{abstract}


\maketitle

\section{Motivation}
Software evolution research often asks why duplicated implementations persist
when a reusable abstraction is available. Educational repositories offer a
useful setting for this question because they contain large numbers of real
artifacts built under constrained programming environments. \scratch is
particularly suitable: code can be duplicated directly, an everyday reuse
mechanism in block-based programming~\citep{robles2017software}, while \cbs provide a
local procedural reuse mechanism. Because a \cb is scoped to a sprite, we
focus on intra-sprite clones, where procedural abstraction is structurally
available, and exclude inter-sprite duplication caused by copying sprites.

We treat as a working premise, to be tested rather than assumed, that
duplication here is an unrealised abstraction opportunity rather than a
deliberate design choice~\citep{kapser2008cloning}. The full
study~\citep{hidalgo2026cloning} provides the methodological and analytical
detail; here we summarize the evidence relevant to that question.

\section{Study and Findings}
We analyze 2,041,691 public \scratch projects. Clone detection 
normalizes literal values while preserving block opcodes and input arity, 
then identifies repeated six-token sequences~\citep{roy2007survey} and extends them to maximal 
clone instances. \cbs are extracted from project JSON together with their 
definitions, parameters, and calls. A \emph{structural match} is recorded when 
a normalized clone sequence also
occurs as a \cb implementation in the same project or elsewhere in the corpus.
Project sophistication is represented by the Dr.~Scratch Computational Thinking~(CT) score~\citep{robles2025dr}.

We find 126,794 projects (6.2\%) containing clones and 242,091 (11.9\%)
containing \cbs. The structural matching analysis gives 70,400 same-project
matches (16.1\%) and 77,506 matches found only outside that project
(17.8\%) among clone instances. These values are not a direct estimate
of ``avoidable clones'': some matches cannot be extracted unchanged because
of Scratch event-trigger constraints, while the six-block detection floor
misses shorter duplicated sequences. In a sensitivity analysis, lowering the
threshold increased the number of matched sequence types by roughly 3.7 times.

Among the 100 most frequent clone patterns, 83 have counterparts implemented as
\cbs elsewhere in the corpus, and manual inspection finds 15 of the remaining 17
expressible through custom-block extraction (two are genuinely ambiguous).
Recurrent initialization, control-flow, animation, data, and drawing sequences
thus appear in the corpus both as duplicated code and as procedures.

Many abstractions remain narrow in scope: 53.0\% of \cbs accept no parameters, and 
43.4\% contain only one or two blocks. Encapsulating a sequence in a named unit
is not the same as supporting reuse across contexts through parameterization.

Across CT-score bands, cloning and \cb use increase together rather than in a
substitutive pattern. Clone prevalence rises from 0.40\% in the beginner band
to 48.17\% in the expert band, while \cb adoption rises from 0.54\% to
77.46\%; combined use grows monotonically from 0.04\% of beginner projects
to 42.02\% of expert projects. The association is
cross-sectional and is influenced by remix inheritance and project size, but
its direction is preserved in original-project and size-normalized analyses
(Spearman $\rho = 0.22$; full effect sizes in the paper).
Thus, higher CT-score levels are associated with a broader repertoire of
reuse mechanisms, not with the disappearance of cloning. This coexistence
may also be environmental: \scratch affords duplication directly, but offers
no duplicate detection, extraction, or parameterization support, so persistence
may reflect the tooling as much as learner choice. 

\section{Implications for Software Evolution}
These results argue against reading persistent duplication as a deliberate design
choice, so clones default to refactoring candidates in block-based programming. The corpus shows repeated
structural patterns that are compatible with abstraction, yet abstraction and
duplication coexist even in projects with high CT scores: projects that use procedural
abstraction may still contain substantial code duplication. This raises
an specific question for software evolution research: when is duplicated code worth
abstracting?

This also yields a concrete intervention question. If the absence of refactoring
support contributes to the coexistence we observe, facilities such as duplicate
detection, ``extract to custom block'' refactoring~\citep{fowler2018refactoring}, and parameterization
assistance should reduce the cost of consolidating repeated code. The resulting
question remains to be tested: Does making refactoring easier 
change how code is reused in evolving block-based programs? 

The full paper~\citep{hidalgo2026cloning} discusses the detection method, 
robustness analyses, limitations, and pattern catalogue in detail, and 
includes the replication package.

\end{document}